\documentstyle[11pt,epsfig]{article}
\title{
Valence Quark Polarization In the Nucleon And the Deuteron Data }

\author{Firooz Arash$^{(a)}$\footnote{e-mail: farash@cic.aut.ac.ir} and
Fatemeh Taghavi-Shahri$^{(b)}$\\
$^{(a)}$ Physics Department, Tafresh University, Tafresh, Iran \\
$^{(b)}$
Physics Department, Iran University of Science and Technology, Narmak, Iran \\
}
\date{\today}
\begin{document}
\maketitle
\begin{abstract}
Within the framework of the so-called {\it{valon}} model, We argue
that a substantial part of the nucleon spin, about $40\%$, is
carried by the polarized valence quarks. The remaining is the
result of cancelations between gluon polarization and the orbital
angular momentum, where the gluon polarization is the dominant
one. It is shown that the sea quark contributions to the spin of
any hadron is simply marginal and consistent with zero. Our
findings point to a substantially smaller value for $a_8$ than
inferred from hyperon - $\beta$ decay, suggesting that full
$SU(3)$ symmetric assumption needs to be reconsidered. New and
emerging experimental data tend to support this finding. Finally,
we show that within the model presented here the experimental data
on the polarized structure functions $g_{1}^{p, n, d}$ are
reproduced.
\end{abstract}
\section{INTRODUCTION}
The most direct tool and sensitive test for probing the quark and
gluon substructure of hadrons is the polarized Deep Inelastic
Scattering (DIS) processes. In such experiments detailed
information can be extracted on the shape and magnitude of the
spin dependent parton distributions, $\delta q_{f}(x,Q^{2})$. Deep
inelastic scattering reveals that the nucleon is a rather
complicated object consisting of an infinite number of quarks,
anti-quarks, and gluon. Contributions from these components to the
spin structure of hadron is an ongoing source of debates.
Substantial activities, both in theory and in experiment, are
performed to disentangle various contributions. The analysis of
 \cite{1} reveals that the gluon and the orbital angular momentum
contribute to the spin of nucleon with opposite signs. It turns
out that the contribution from gluon polarization is the dominant
factor, accounting for about $60 \% $ of the spin of proton. This
still leaves a substantial amount of nucleon spin to come from
quark sector: valence and sea. Expectations are that the sea is
localized in the low $x$ region and the valence quark is dominant
at $x\geq 0.3$. The role of sea quark is although still unclear,
but both theoretical \cite{1} and experimental investigations
\cite{2} \cite{3}\cite{4} point in the direction that it should be
marginal. The purpose of this letter is to investigate valence and
sea quark contributions to the nucleon spin. COMPASS collaboration
has recently published accurate data on the spin structure of
deuteron \cite{4} and has measured semi-inclusive difference
asymmetry, $A^{h^{+} -h^{-}}$, for hadrons of opposite charge in
the kinematic region $0.006< x< 0.7$. These measurements determine
the valence polarization and allows for the evaluation of the
first moment of $\Delta u_{v}+\Delta d_{v}$. COMPASS coll. have
used a hybrid procedure, in that they have taken the MRST04
leading order parametrization of the unpolarized parton
distributions along with the leading order fit of DNS polarized
parton distributions \cite{5} to evolve values of $\Delta
u_{v}+\Delta d_{v}$ to a common $Q^2$ value fixed at $10$
$(GeV/c)^2$. The message of this experiment is threefold. First,
they provide a set of accurate data on $g_{1}^{d}$, which is
valuable in its own right. Secondly, they imply that the total sea
quark contributions to the spin of nucleon is consistent with
zero. This finding is in agreement with the earlier results from
HERMES collaboration \cite{2} \cite{3} thus, reinforcing the point
further. Neither group, however, provide an explanation for their
findings. Thirdly, it seems that their results can accommodate
both positive and negative gluon polarization, $\Delta G.$ \\
It is important to investigate the implications of the new set of
COMPASS data in an attempt to offer some explanations for these
findings. The majority of the theoretical studies have been
focused on the singlet component of the polarized structure
function in order to explain its smallness. This has led to the
assumption that the non-singlet component is rather well
understood. Experimental evidence for this assumption comes from
the confirmation of the Bjorken sum rule which relates the first
moments of $g_{1}^{p}$ and $g_{1}^{n}$. However, the Bjorken sum
rule depends merely on the fundamental $SU(2)$ isospin symmetry
between the matrix elements of charged and neutral axial currents
and therefore expected to hold, does not entirely fix the first
moment of non-singlet component of $g_{1}^{p}$. This component, in
the leading order, depends on $a_{8}$ which is under the
assumption of $SU(3)_{f}$ symmetry usually taken to be
$3F-D=0.579\pm0.025$. Information that can be obtained from
baryonic $\beta$ decay is highly limited and only provides
information about the first moment of non-singlet part of
$g_{1}^{p}$. As a result, when it is used to extract sea quark
contributions to the proton spin, the outcome becomes less
reliable. In the model described in the next section, we will show
that the calculations in the framework of NLO perturbation theory,
shows that the sea quark contribution to the spin of proton is
essentially consistent with zero and the value for $a_{8}$ is
substantially different from $0.579\pm0.025$, a result that seems
confirmed by the emerging experiments.
\section{The Model}
As in \cite{1}, our approach in addressing the above issues is
based on the so-called {\it{valon model}}. In this model a nucleon
is composed of three dressed valence quarks, the valons. Each
valon has its own internal structure which can be probed at high
enough $Q^2$. At low $Q^2$, a valon behaves as a valence quark.
The internal structure of a valon is calculated in the
Next-to-Leading Order in QCD. We have worked in $\overline{MS}$
scheme with $\Lambda_{QCD}=0.22$ GeV and $Q_{0}^{2}=0.283$
$GeV^2$. The details are given in \cite{1}. It turns out that in a
polarized valon the sea quark polarization is consistent with
zero, whereas the valence quark carries almost entire spin of the
valon. In fact, as it is evident from Figure 1, for $Q^2\geq 1$
$GeV^2$ the value of $\delta \Sigma_{valon}\simeq 0.88$ and
remains so, almost independent of $Q^2$. Between $Q_{0}^{2}=
0.283$ $GeV^{2}$ and $\sim 1$ $GeV^{2}$ it decreases from unity to
$0.88$. For the same range the value of $\delta q_{sea}$ for
individual flavor is around $-0.002$ and hence, marginal. The fact
that sea quark polarization in the valon is consistent with zero
can be understood on the theoretical grounds. The valon structure
is generated by perturbative dressing in QCD. In such processes
with massless quarks, helicity is conserved and therefore, the
hard gluons cannot induce sea polarization perturbatively.
\begin{figure}
 \epsfig{figure=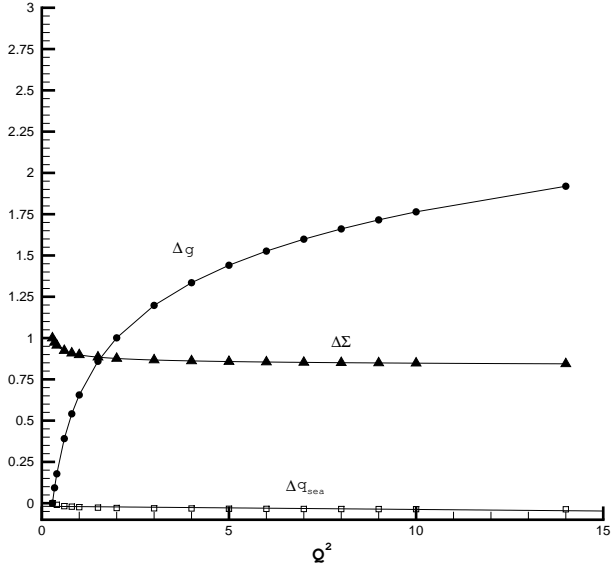,width=10cm}
\caption{\footnotesize First Moments, $\Delta g(n=1, Q^2)$,
$\Delta \Sigma(n=1, Q^2)$, and
 $\Delta q_{sea}(n=1, Q^2)$ of various components in a valon as a function of $Q^2$ }
\label{figure 1.}
\end{figure}
So, it appears that $\delta \Sigma_{valon}$ is very close to
unity. However, the gluon polarization in a valon is positive and
increases with $Q^2$, making the spin picture of a valon rather
bluer. Spin of a valon is $\frac{1}{2}$ and the following sum rule
applies.
\begin{equation}
\frac{1}{2}=\frac{1}{2}\Delta \Sigma + \Delta g +L_{z}.
\end{equation}
This picture shows that the gluon polarization in a valon is
almost entirely compensated by the negative orbital angular
momentum contribution. We would like to make it clear that in this
work we have not attempted to solve evolution equation for
$L_{z}$; it is determined simply by evaluating other quantities in
Eq.(1) and then solving it for $L_{z}$. In Figure 2, we present
the total orbital angular momentum of the partons in a valon. P.G.
Ratcliffe \cite{6} was the first to argue that the orbital angular
momentum has a negative contribution to the spin of proton. In a
remarkable paper by Ji, Tang, and hoodbhoy \cite{7} it is shown
that in the nucleon both the orbital angular momentum and the
helicity of gluon grow with opposite signs. It is also shown that
the total orbital angular momentum carried by quark and gluon is
negative; a conclusion that we have also arrived at. In a recent
paper \cite{8} Anthony Thomas has also concluded that based on the
standard features of the non-perturbative structure of the
nucleon, the majority of the proton spin comes from the orbital
angular momentum of $u$ and $\overline{u}$ quarks.
\begin{figure}
 \epsfig{figure=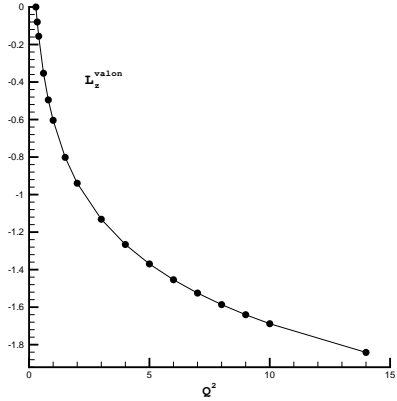,width=8cm}
\caption{\footnotesize Total orbital angular momentum of partons,
$L_{z}^{valon}(Q^2)$, in a valon as a function of $Q^2$. }
\label{figure 2.}
\end{figure}
Having specified the various components that contribute to the
spin of a valon, we now turn to the hadron structure, which is
obtained by a convolution integral as follows
\begin{equation}
g^{h}_{1}(x,Q^{2})=\sum_{\it{valon}}\int_{x}^{1}\frac{dy}{y}
\delta G_{\it{valon}}^{h}(y) g^{\it{valon}}_{1}(\frac{x}{y},Q^{2})
\end{equation}
where $\delta G_{\it{valon}}^{h}(y)$ is the helicity of the valon
in the hosting hadron and $g^{\it{valon}}_{1}(\frac{x}{y}, Q^{2})$
is the polarized structure function of the valon. The latter, say,
for a U-type valon, is simply given by
\begin{equation}
2 g_{1}^{U}(z,Q^{2})=\frac{4}{9}(\delta G_{\frac{u}{U}}+ \delta
G_{\frac{\bar{u}}{U}}) +\frac{1}{9}(\delta G_{\frac{d}{U}}+\delta
G_{\frac{\bar{d}}{U}}+\delta G_{\frac{s}{U}}+\delta
G_{\frac{\bar{s}}{U}}) +...
\end{equation}
$\delta G_{\it{valon}}^{h}(y)$, is obtained  from unpolarized
valon distribution by
\begin{equation}
\delta G_{j}(y) = \delta F_{j}(y) G_{j}(y)
\end{equation}
where, $G_{j}(y)$, are the unpolarized valon distributions and are
determined by a phenomenological argument for a number of
hadrons\cite{9} \cite{10} \cite{11}. The functions $\delta
F_{j}(y)$ ,where $j$ refers to U and D type valons, are given in
\cite{1}. In Figure 3 the shape of $\delta G_{U,D}(y)$ are shown.
\begin{figure}
 \epsfig{figure=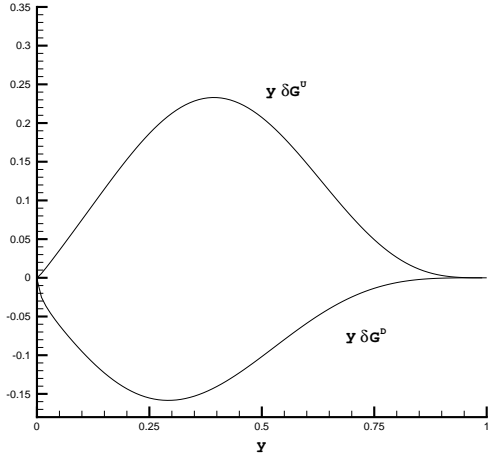,width=8cm}
\caption{\footnotesize Polarized valon distributions,$y\Delta
G_{U,D}(y)$ in proton } \label{figure 3.}
\end{figure}
The helicity distributions of various parton types in a hadron is
obtained by
\begin{equation}
\delta q_{\frac{i}{h}}(x,Q^2)=\sum \int_{x}^{1} \frac{dy}{y}\delta
G_{\frac{valon}{h}}(y) \delta q_{\frac{i}{valon}}(\frac{x}{y},Q^2)
\end{equation}
Here It suffice to give the parametric form of these helicity
distribution functions in the proton
\begin{equation}
x\delta q_{\frac{i}{p}}(x,Q^2)=a_{i}x^{b_i}(1-x)^{c_i}
(1+d_{i}x^{0.5}+e_{i} x+f_{i} x^2)
\end{equation}
where $a_{i},b_{i},. . . $ are functions of $Q^2$ and the index
$i$ refers  to $u_{valence}$, $d_{valence}$, sea quark, and gluon.
\section{Comparison with  the experimental data}
We are now in a position to consider the experimental data at the
hadron level. Substituting Eqs. (3) and (4) into equation (2)
gives the polarized structure of  the proton and  the neutron. The
polarized structure function of deuteron is obtained by evaluating
\begin{equation}
g_{1}^{d}=\frac{1}{2}(g_{1}^{p}+g_{1}^{n})(1-1.5 \omega_D).
\end{equation}
We have calculated $g_{1}^{d}$ at $Q^{2}=10$ $(Gev/c)^2$ and
$Q^{2}=3$ $(Gev/c)^2$ and compared our results with the
experimental data from \cite{4} along with the other global fit
results in Figure 4.
\begin{figure}
 \epsfig{figure=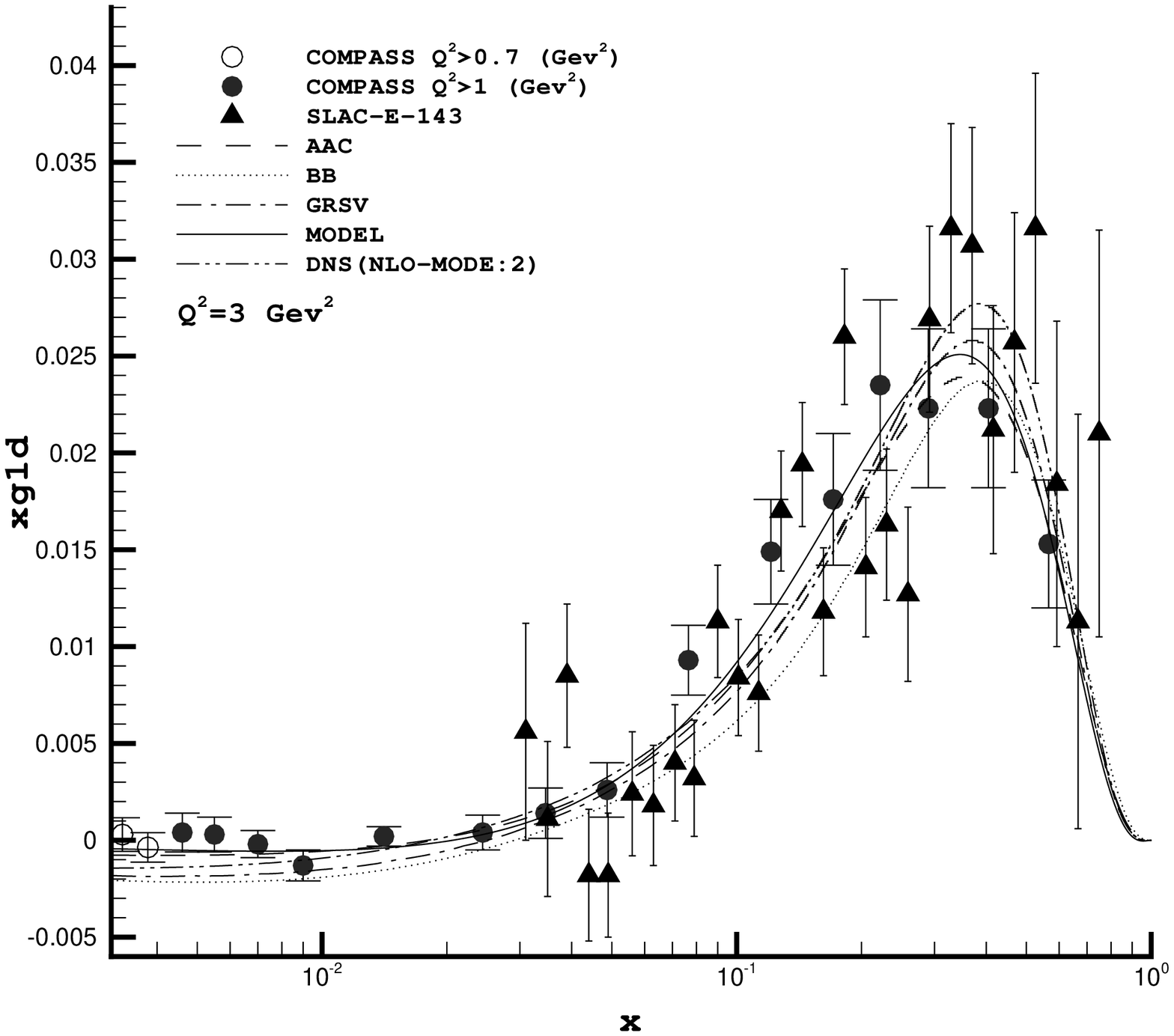,width=7cm}
 \epsfig{figure=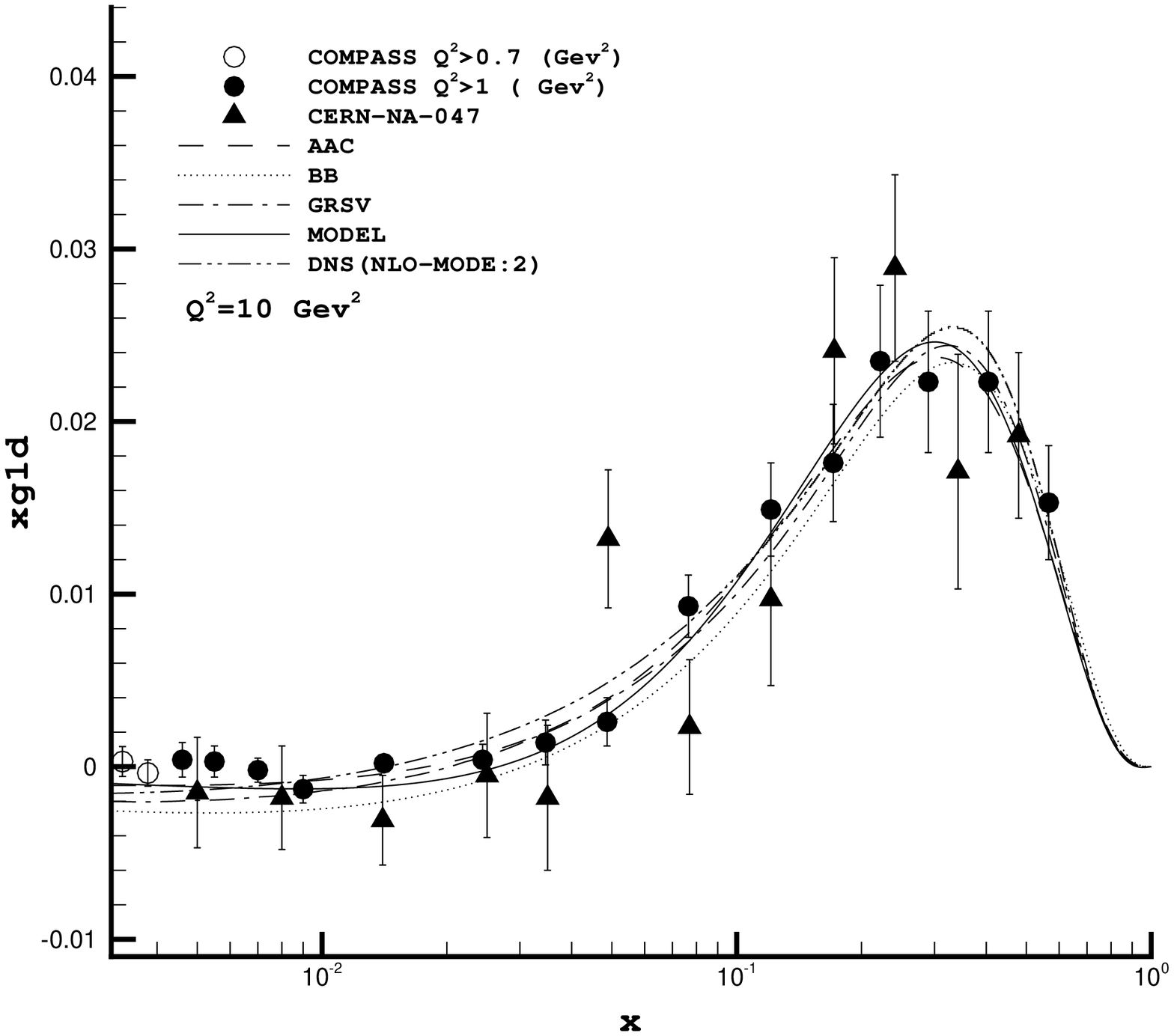,width=7cm}
\caption{\footnotesize Deuteron structure function $g_{1}^{d}$ at
$Q^2=3$ and $Q^2=10$ $GeV^2$. For the purpose of comparison
results from global fits are also shown. } \label{figure 4.}
\end{figure}
In contrast to DNS, AAC, and BB fits, our results are obtained
without any fit to the experimental data. As stated earlier, the
sea polarization in a valon is very small and hence, more so at
the hadron level, being consistent with zero. This finding is
interesting and is in excellent agreement with earlier results
from HERMES collaboration \cite{2} \cite{3}. COMPASS collaboration
although does not provide direct information about the sea
polarization, but clearly indicates that their data is consistent
with $\Delta \overline{u}_{sea}+\Delta \overline{d}_{sea}\sim 0$
and suggests that $\Delta \overline{u}_{sea}$ and $\Delta
\overline{d}_{sea}$, if different from zero, must be of opposite
sign \cite{12}. Forthcoming COMPASS data on proton target will
provide separate determinations of $\Delta \overline{u}_{sea}$ and
$\Delta \overline{d}_{sea}$. The reconfirmation of HERMES findings
on sea polarization as well as our results, naturally shifts the
attention to the valence and gluon polarizations. The situation
with the strange sea quark polarization is less clear because, the
information about $\Delta s$ and $\Delta \overline{s}$ is obtained
mainly from hyperon $\beta$ decay via the value of
$a_{8}=3F-D=0.579\pm 0.025$ or in models like the
Nambu-Jona-Lasinio models. The way that $a_{8}$ is determined is
based on the full $SU(3)$ symmetry between hyperon decay matrix
elements of the flavor-changing weak axial and neutral currents.
However, there are serious objections to this approach
\cite{13}\cite{14}\cite{15}. The results obtained from
Nambu-Jona-Lasinio model also is not free from ambiguities \cite
{16}. Our findings for $a_{8}$ is $0.39$. This value is
substantially smaller that the value inferred from hyperon data,
but there are already experimental evidences in support of it.
HERMES collaboration \cite{17} gives the value $0.274\pm 0.026\pm
0.011$ in the measured range of $0.02<x<0.6$. The newer and more
accurate data from COMPASS coll. which is extended the measured
range to $0.006<x<0.7$ disfavors the assumption of a flavor
symmetric polarized sea at a two $\sigma$ level and sets the value
of $a_{8}$ at $0.40\pm 0.07\pm 0.05$. A detailed analysis of
non-singlet structure functions in the next-to-leading order is
given in \cite{18} where the assumption of $SU(3)$ symmetry is
studied and interesting conclusions about $a_{8}$ and $a_3$ are
arrived. In contrast to $a_{8}$, our results in \cite{1} for the
Bjorken sum rule, or equivalently, $a_3$, is rather accurate and
agrees with the accepted value $a_{3}=F+D=g_{A}=1.2573\pm 0.0028$. \\
Accepting HERMES results \cite{17}\cite{19} and the data from
COMPASS collaboration [4,12] points to the fact that the role of
sea polarization is marginal and consistent with zero. This fact
is naturally explained in our model, which relies on the
Next-to-Leading order calculations in perturbative QCD. The HERMES
Collaboration has just published \cite{20} a new set of data,
based on the charged-kaon production in deep inelastic scattering
on the deuteron. Their new findings and analysis is essentially a
reconfirmation of our results; that is, strange sea quark helicity
distribution is zero within the experimental uncertainties
\cite{20}. Furthermore, they find that $a_{8}=0.285\pm 0.046\pm
0.057$. This situation leaves us with the valence and gluon
polarization and the orbital angular momentum in the nucleon. The
gluon and orbital angular momentum contributions to the spin of
proton is discussed in \cite{1}. Here our focus is on the valence
sector. Our findings are shown in Figures 5 at $Q^2=10$ $GeV^2$.
Our results for $\Delta \Sigma$, the total quark contribution to
the spin of proton, with $\Delta q_{sea}\sim 0$, lies in the range
of $0.410-0.420$ for $Q^2=[1,10]$ $GeV^2$. The variation of
$\Delta \Sigma$ is due to (marginal) $Q^2$ dependence of $\Delta
q_{v}$ in the Next-to-Leading Order; since $\Delta
P^{(1)}_{NS}\neq 0$. This value of $\Delta \Sigma$ is in excellent
agreement with the results from COMPASS experiments \cite{12}
which gives
\begin{equation}
\Gamma_v=\int_{0}^{1}dx (\Delta u_{v}(x)+\Delta d_{v}(x))=0.41\pm
0.07 \pm 0.05
\end{equation}
and with the results of HERMES \cite{20} which is $0.359\pm 0.026
\pm 0.018$.
\begin{figure}
\centerline{\begin{tabular}{cccc} \epsfig{figure=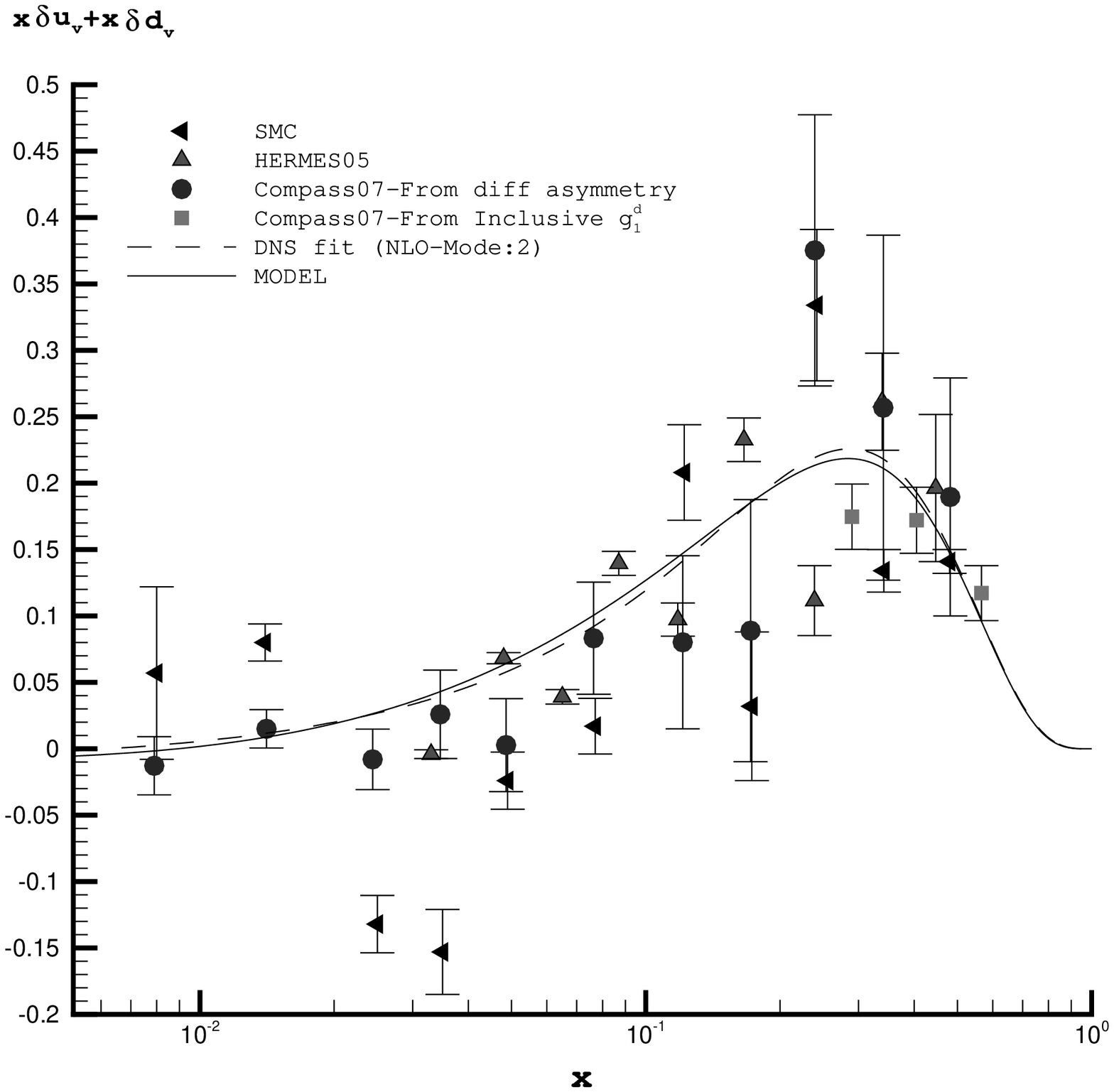,width=8cm}
\epsfig{figure=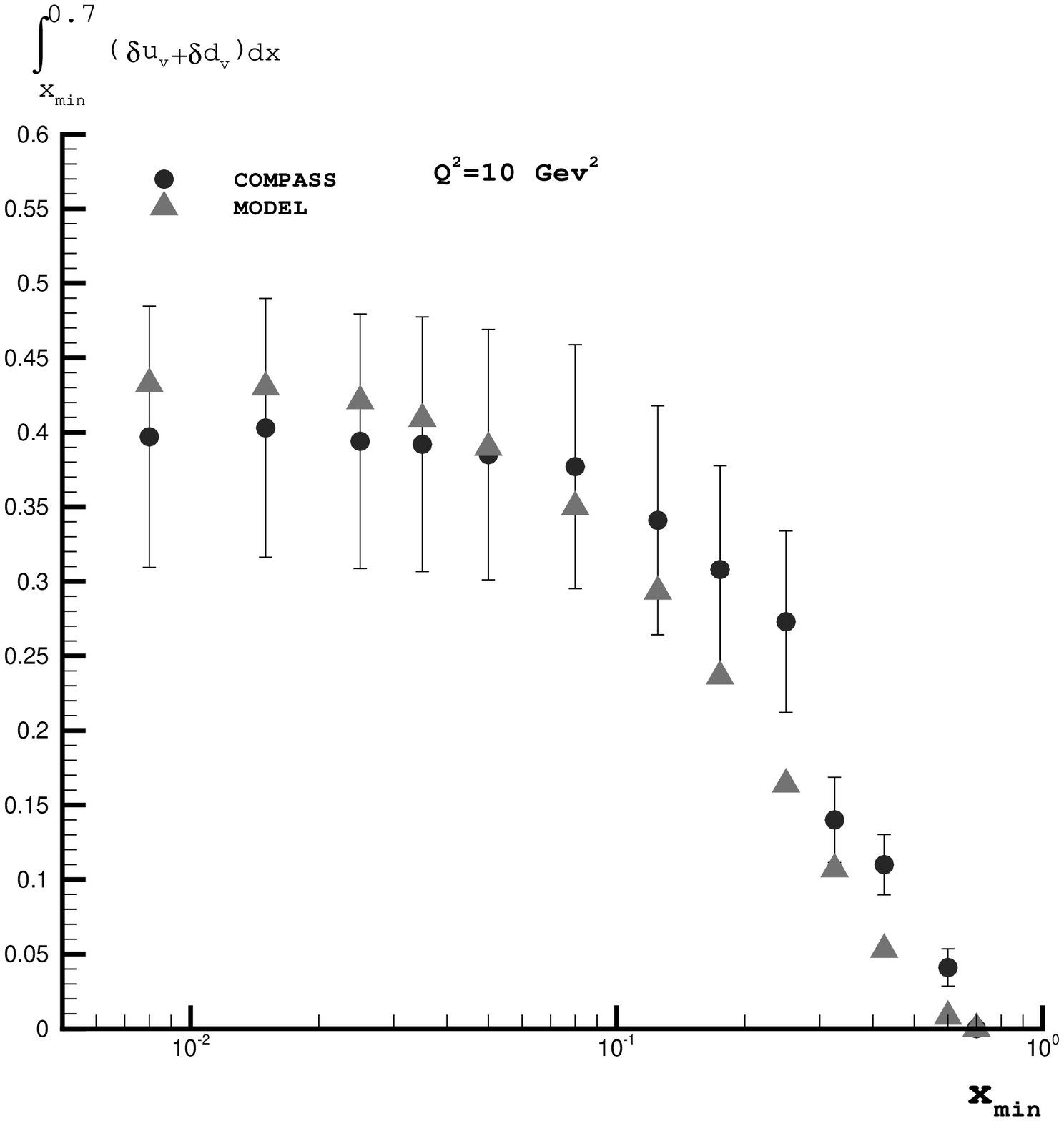,width=8cm}
\end{tabular}}
\caption{\footnotesize {\it{Left:}} $x\delta u_v(x)+x\delta d_v
(x)$ at $Q^2=10$ $(\frac{GeV}{c})^2$.The data points are from
\cite{SMC} \cite{2} \cite{3} \cite{10}. Full line curve is the
model results and the dashed curve in DNS next-to-leading order
fit results. {\it{Right:}} The integral of $\Delta u_v(x)+\Delta
d_v(x)$ over the range $0.006 < x < 0.7$ measured by COMPASS coll.
as the function of the low $x$ limit of integration,$x_{min}$
evaluated at $Q^2=10$ $(\frac{GeV}{c})^2$.The results from the
model calculation are compared with the experimental data.}
\label{figure 5.}
\end{figure}
\section{Conclusions}
We have shown that the valon representation is a suitable
phenomenological frame work to deal with the polarization
structure functions of hadrons. Within this representation a valon
is a dressed quark and its spin arises from an interplay among the
spin of the valence quark, the gluon polarization and the orbital
angular momentum components. It also shows that the sea quark
polarization in a valon is marginal. A negative orbital angular
momentum emerges and competes with the gluon polarization,
resulting in the partial cancelation of the later. There is no net
sea polarization at the hadron level; a result that agrees rather
well with the existing experimental data. The valence quark
contribution to the spin of proton stands at around $40\%$ and the
remaining comes from the gluon polarization. We have further shown
that the results of this model predicts a substantially smaller
value for $a_8$ than inferred from the hyperon $\beta$ decay.
Experimental evidences for this finding is also gradually
emerging, suggesting that the assumption of full $SU(3)$ symmetric
sea needs reassessments. Finally, we have demonstrated that the
model accommodates the wealth of experimental data on the
polarized structure functions, $g_{1}^{p}, g_{1}^{n}, g_{1}^{d}$
rather nicely.

\end{document}